\documentclass[pra,aps,superscriptaddress,showpacs,floatfix,tightenlines, twocolumn]{revtex4-1}
\usepackage{amsfonts}
\usepackage{graphicx}
\usepackage{bm}
\usepackage{amssymb}
\usepackage{amsmath}

\begin{document}

\title{Remedying the strong monotonicity of the coherence measure in terms
of the Tsallis relative $\alpha$ entropy }
\author{Haiqing Zhao}
\affiliation{College of Science, Dalian jiaotong University, Dalian, 116028,
China}
\author{Chang-shui Yu}
\email{ycs@dlut.edu.cn}
\affiliation{School of Physics and Optoelectronic Technology, \\
Dalian University of Technology, Dalian 116024, China}
\date{\today }

\begin{abstract}
Coherence is the most fundamental quantum feature of the nonclassical
systems. The understanding of coherence within the resource theory has been
attracting increasing interest among which the quantification of coherence
is an essential ingredient. A satisfactory measure should meet certain
standard criteria. It seems that the most crucial criterion should be the
strong monotonicity, that is, average coherence doesn't increase under the
(sub-selective) incoherent operations. Recently,  the Tsallis relative $\alpha $
entropy [A. E. Rastegin, Phys. Rev. A \textbf{93}, 032136 (2016)] has been tried to quantify the coherence. But it was shown to violate the strong monotonicity, even though it can unambiguously
distinguish the coherent and the incoherent states with the monotonicity. 
Here we establish a family of coherence quantifiers which are closely related to the
Tsallis relative $\alpha $ entropy. It proves that this family of quantifiers satisfy all the standard
criteria and particularly cover several typical coherence measures. \end{abstract}

\maketitle

\section{Introduction}

Coherence, the most fundamental quantum feature of a nonclassical system,
stems from quantum superposition principle which reveals the wave particle
duality of matter. It has been shown that coherence plays the key roles in
the physical dynamics in biology  \cite{Engel,Plenio,Coll,loyd,licm,Huel}, transport theory \cite{reb,wit} , and thermodynamics \cite{berg,Nar,Horo,Los1,Los2}. In
particular, some typical approaches such as phase space distributions and
higher order correlation functions have been developed in quantum optics to
reveal quantum coherence even as an irrigorous quantification \cite{Glauber,Sudarshan,Scully}. Quite
recently, quantum coherence has been attracting increasing
interest in various aspects \cite%
{Pleniom, Giro, Napoli, Lewen,Rast,Piani,Winter,Du,Chi,Chi2,Chi3,Marvian,Marvian2,Yao,Sing,Radha} including the quantification of coherence  \cite{Pleniom, Giro, Napoli, Lewen,Rast,Piani}, the operational resource theory \cite{Winter, Du, Chi,Chi2,Chi3}, the distribution \cite{Radha}, the different
understandings \cite{Yu09,Stre,Ma,Tan} and so on.

Quantification of coherence is the most essential ingredient not only in the
quantum theory but also in the practical application. Various quantities
have been proposed to serve as a coherence quantifier, however the available candidates are still quite limited. Up to now, only two
alternatives, i.e., the coherence measures based on $l_{1}$ norm and the relative entropy, have turned
out to be a satisfactory coherence measure \cite{Pleniom}. In contrast, the  usual $l_{p}$ ($p\neq 1$) norm can not directly induce a good
measure \cite{Lewen}. In addition, the coherence quantifier based on the Fidelity is easily shown to
satisfy the monotonicity that the coherence of the post-incoherent-operation state doesn't increase, 
but it violates the strong monotonicity that average coherence doesn't increase under the
sub-selective incoherent operations \cite{Pleniom,Fan}. Similarly, even though the coherence based on the trace norm
 also satisfies the monotonicity but lacks a strict proof for the strong monotonicity \cite{Lewen,Fan}. However, we know
that the strong monotonicity is much more important than the monotonicity
not only because the sub-selection of the measurement outcomes required by
the strong monotonicity can be well controlled in experiment as is stated in
Ref. \cite{Pleniom,Lewen}, but also because the realizable sub-selection would  lead to the real
increment of the coherence from the point of resource theory of view if the
strong monotonicity was violated. In this sense, the quantitative
characterization of coherence still needs to be paid more attention.

Recently, Ref. \cite{Rast} has also proposed a coherence quantifier in terms of the
Tsallis relative $\alpha $ entropy which lays the foundation to the
non-extensive thermo-statistics and plays the same role as the standard
logarithmic entropy does in the information theory \cite{lisa,Tsallis}. However, it is
unfortunate that the Tsallis relative $\alpha $ entropy isn't an ideal
coherence measure either because Ref. \cite{Rast} showed that it only satisfies the
monotonicity and a variational monotonicity rather than the strong
monotonicity. Is it possible to bridge the Tsallis relative $%
\alpha $ entropy with the strong monotonicity by some particular and
elaborate design? In this paper, we build such a bridge between the Tsallis relative $\alpha $ entropy with
the strong monotonicity, hence present a family of good coherence
quantifiers. By considering the special case in this family, one can find that  the $l_{2}$ norm
can be validly employed to quantify the coherence. The remaining of this
paper is organized as follows. In Sec. II, we introduce the coherence measure and  the Tsallis
relative $\alpha $ entropy. In Sec. III, we present the
family of coherence quantifier and mainly prove them to be strongly
monotonic.  In Sec. IV, we study the maximal coherence and several particular coherence measure. Finally, we finish the paper by
the conclusion and some discussions.

\section{The coherence and the Tsallis relative $\protect%
\alpha $ entropy}

The resource theory includes three ingredients: the free states, the
resource states and the free operations \cite{Winter, Gour}. For coherence, the free states are
referred as to the incoherent states which are defined in a given fixed
basis $\left\{ \left\vert i\right\rangle \right\} $ by the states with the
density matrices in the diagonal form, i.e., $\delta =\sum\limits_{i}\delta
_{i}\left\vert i\right\rangle \left\langle i\right\vert $ with $%
\sum\limits_{i}\delta _{i}=1$ for the positive $\delta _{i}$. All the states
without the above diagonal form are the coherent states, i.e., the resource
states. The quantum operations described by the Kraus operators $\left\{
K_{n}\right\} $ with $K_{n}^{\dag }K_{n}=\mathbf{I}$ are called as the
incoherent operations and serve as the free operations for coherence, if $%
K_{n}\delta K_{n}^{\dag }\in \mathcal{I}$ for any incoherent $\delta $. In
this sense, the standard criteria of a good coherence quantifier $C(\rho )$
for the state $\rho $ can be rigorously rewritten as \cite{Pleniom} (i) (Null) $C(\delta
)=0 $ for $\delta \in \mathcal{I}$; (ii) (Strong monotonicity) for any state 
$\rho $ and incoherent operations $\left\{ K_{n}\right\} $, $C(\rho
)\geqslant \sum\limits_{n}p_{n}C(\rho _{n})$ with $p_{n}=$Tr$K_{n}\rho
K_{n}^{\dag }$ and $\rho _{n}=K_{n}\rho K_{n}^{\dag }/p_{n}$; (iii)
(Convexity) For any ensemble $\left\{ q_{i},\sigma _{i}\right\} $, $%
C(\sum\limits_{i}q_{i}\sigma _{i})\leq \sum\limits_{i}q_{i}C(\sigma _{i})$.
In addition, the monotonicity requires $C(\rho )\geqslant
C(\sum\limits_{n}p_{n}\rho _{n})$, however, it alone isn't laid in an
important position because the measurement outcomes of $\left\{
K_{n}\right\} $ can be well controlled (subselected) in practical
experiments, or in other words, the violation of the strong monotonicity
means that the ultimate coherence is actually increased by the incoherent
operations $\left\{ K_{n}\right\} $ even though it can be automatically
implied by (ii) and (iii). With these criteria, any measure of
distinguishability such as the (pseudo-) distance norm could induce a
potential candidate for a coherence quantifier. But it has been shown that
some candidates only satisfy the monotonicity rather than the strong
monotonicity, so they are not ideal and could be only used in the limited
cases. Ref. \cite{Rast} found that the coherence based on the Tsallis relative $%
\alpha $ entropy is also such a coherence quantifier without the strong
monotonicity.

The Tsallis relative $\alpha $ entropy is a special case of the quantum $f$%
-divergences \cite{Rast,Hiai}. For two density matrices $\rho $ and $\sigma $, it is defined
as%
\begin{equation}
D_{\alpha }\left( \rho ||\sigma \right) =\frac{1}{\alpha -1}\left( \text{Tr}%
\rho ^{\alpha }\sigma ^{1-\alpha }-1\right)
\end{equation}%
for $\alpha \in (0,2]$. It is shown that for $\alpha \longrightarrow 1,$ $%
D_{\alpha }\left( \rho ||\sigma \right) $ will reduce to the relative
entropy $S\left( \rho ||\sigma \right) =Tr\rho \log _{2}\rho -\rho \log
_{2}\sigma $. The Tsallis relative $\alpha $ entropy $D_{\alpha }\left( \rho
||\sigma \right) $ inherits many important properties of the quantum $f$%
-divergences, for example, (Positivity) $D_{\alpha }\left( \rho ||\sigma
\right) \geq 0$ with equality if and only if $\rho =\sigma $, (Isometry) $%
D_{\alpha }\left( U\rho U^{\dag }||U\sigma U^{\dag }\right) =D_{\alpha
}\left( \rho ||\sigma \right) $ for any unitary operations, (Contractibility) $%
D_{\alpha }\left( \$\left( \rho \right) ||\$(\sigma )\right) \leq D_{\alpha
}\left( \rho ||\sigma \right) $ under any trace-preserving and completely
positive (TPCP) map $\$$ and (Joint convexity) $D_{\alpha }\left(
\sum_{n}p_{n}\rho _{n}||\sum_{n}p_{n}\sigma _{n}\right) \leq
\sum_{n}p_{n}D_{\alpha }\left( \rho _{n}||\sigma _{n}\right) $ for the
density matrices $\rho _{n}$ and $\sigma _{n}$ and the corresponding
probability distribution $p_{n}$.

Based on the Tsallis relative $\alpha $ entropy $D_{\alpha }\left( \rho
||\sigma \right) $, the coherence in the fixed reference basis $\left\{
\left\vert j\right\rangle \right\} $ can be characterized by \cite{Rast}
\begin{eqnarray}
\tilde{C}_{\alpha }(\rho ) &=&\min_{\delta \in \mathcal{I}}D_{\alpha }\left(
\rho ||\delta \right)   \notag \\
&=&\frac{1}{\alpha -1}\left[ \left( \sum_{j}\left\langle j\right\vert \rho
^{\alpha }\left\vert j\right\rangle ^{1/\alpha }\right) ^{\alpha }-1\right] .
\label{Tsc}
\end{eqnarray}%
However, it is shown that $\tilde{C}_{\alpha }(\rho )$ satisfies all the
criteria for a good coherence measure but the strong monotonicity. Since $%
D_{\alpha \rightarrow 1}\left( \rho ||\sigma \right) $ reduces to the
relative entropy $S\left( \rho ||\sigma \right) $ which has induced the good
coherence measure, throughout the paper we are mainly interested in $\alpha
\in (0,1)\cup (1,2]$.

In addition, the Tsallis relative $\alpha $ entropy $D_{\alpha }\left( \rho
||\sigma \right) $ can also be reformulated by a very useful function as
 \begin{equation}
D_{\alpha }\left( \rho ||\sigma \right) =\frac{1}{\alpha -1}\left( f_{\alpha
}\left( \rho ,\sigma \right) -1\right) 
\end{equation} 
with 
\begin{equation} 
f_{\alpha }\left( \rho
,\sigma \right) =\mathrm{Tr}\rho ^{\alpha }\sigma ^{1-\alpha }.
\end{equation} 
Accordingly, the
coherence $\tilde{C}_{\alpha }(\rho )$ can also be rewritten as 
\begin{equation}
\tilde{C}%
_{\alpha }(\rho )=\frac{1}{\alpha -1}\left[ \mathrm{sgn}_{1}(\alpha
)\min_{\delta \in \mathcal{I}}\mathrm{sgn}_{1}(\alpha )f_{\alpha }\left( \rho
,\delta \right) -1\right] 
\end{equation} which, based on Eq. (\ref{Tsc}), leads to the
conclusion%
\begin{equation}
\min_{\delta \in \mathcal{I}}\text{sgn}_{1}(\alpha )f_{\alpha }\left( \rho
,\delta \right) =\left( \sum_{j}\left\langle j\right\vert \rho ^{\alpha
}\left\vert j\right\rangle ^{1/\alpha }\right) ^{\alpha }.  \label{concl}
\end{equation}%
Based on Eq. (\ref{concl}) and the properties of $D_{\alpha }\left( \rho ||\sigma \right) $ mentioned above, one can have the following observations for the function $f_{\alpha }\left( \rho ,\sigma \right) $ \cite{Rast, Hiai}.

\textbf{Observations}:  $f_{\alpha }\left( \rho ,\sigma \right) $ satisfies the following properties:

(I) $f_{\alpha }\left( \rho ,\sigma \right) \geq 1$ for $\alpha \in (1,2]$
and $f_{\alpha }\left( \rho ,\sigma \right) \leq 1$ for $\alpha \in (0,1)$
with equality if and only if $\rho =\sigma $;

(II) For a unitary operation $U$, $f_{\alpha }\left( U\rho U^{\dagger
},U\sigma U^{\dagger }\right) =f_{\alpha }\left( \rho ,\sigma \right) $;

(III) For any TPCP map $\$$, $f_{\alpha }\left( \rho ,\sigma \right) $
doesn't decrease for $\alpha \in (0,1)$, and doesn't increased for $\alpha
\in (1,2]$, namely, 
\begin{equation}
\text{sgn}_{1}(\alpha )f_{\alpha }\left( \$\left[ \rho \right] ,\$\left[
\sigma \right] \right) \leq \text{sgn}_{1}(\alpha )f_{\alpha }\left( \rho
,\sigma \right) ,
\end{equation}%
where the function is defined by sgn$_{1}(\alpha )=\left\{ 
\begin{array}{cc}
-1, & \alpha \in (0,1) \\ 
1, & \alpha \in (1,2]%
\end{array}%
\right. $;

(IV) The function sgn$_{1}(\alpha )f_{\alpha }\left( \rho ,\sigma \right) $
is jointly convex;

(V) For a state $\delta $, $f_{\alpha }\left( \rho \otimes \delta ,\sigma
\otimes \delta \right) =f_{\alpha }\left( \rho ||\sigma \right) $, which can
be easily found from the function itself.

\section{The coherence measures based on the Tsallis relative $\protect%
\alpha $ entropy}

To proceed, we would like to present a very important lemma for the function 
$f_{\alpha }\left( \rho ,\sigma \right) $, which is the key to show our main
result.

\textbf{Lemma 1}.-Suppose both $\rho $ and $\sigma $ simultaneously undergo
a TPCP map $\$:=\left\{ M_{n}:\sum\limits_{n}{M}_{n}^{\dagger }{M}%
_{n}=\mathbb{I}_{S}\right\} $ which transforms the states $\rho $ and $%
\sigma $ into the ensemble  $\left\{ p_{n},\rho _{n}\right\} $ and $\left\{
q_{n},\sigma _{n}\right\} $, respectively, then we have 
\begin{equation}
\text{sgn}_{1}(\alpha )f_{\alpha }\left( \rho _{S},\delta _{S}\right) \geq 
\text{sgn}_{1}(\alpha )\sum\limits_{n}p_{n}^{\alpha }q_{n}^{1-\alpha
}f_{\alpha }\left( \rho _{n},\sigma _{n}\right) .  \label{lem1}
\end{equation}%
\textbf{Proof. }Any TPCP map can be realized by a unitary operation on a
composite system followed by a local projective measurement \cite{Nielsen}. Suppose system
S is of our interest and A is an auxiliary system. For a TPCP map $%
\$:=\left\{ M_{n}:\sum\limits_{n}{M}_{n}^{\dagger }{M}_{n}=\mathbb{%
I}_{S}\right\} $, one can always find a unitary operation $U_{SA}$ and a
group of projectors $\left\{ \Pi _{n}^{A}=\left\vert n\right\rangle
_{A}\left\langle n\right\vert \right\} $ such that 
\begin{eqnarray}
&&M_{n}\rho _{S}M_{n}^{\dagger }\otimes \Pi _{n}^{A}  \notag \\
&=&\left( \mathbb{I}_{S}\otimes \Pi _{n}^{A}\right) U_{SA}\left( \rho
_{S}\otimes \Pi _{0}^{A}\right) U_{SA}^{\dag }\left( \mathbb{I}_{S}\otimes
\Pi _{n}^{A}\right) .  \label{eq}
\end{eqnarray}%
Using Properties (I) and (II), we have 
\begin{eqnarray}
&&f_{\alpha }\left( \rho _{S},\delta _{S}\right)   \notag \\
&=&f_{\alpha }\left( U_{SA}\left( \rho _{S}\otimes \Pi _{0}^{A}\right)
U_{SA}^{\dag },U_{SA}\left( \sigma _{S}\otimes \Pi _{0}^{A}\right)
U_{SA}^{\dag }\right) 
\end{eqnarray}%
holds for any two states $\rho _{S}$ and $\sigma _{S}$. Let $\rho _{Sf}=\$_{SA}%
\left[ U_{SA}\left( \rho _{S}\otimes \Pi _{0}^{A}\right) U_{SA}^{\dag }%
\right] $ and $\sigma _{Sf}=\$_{SA}\left[ U_{SA}\left( \sigma _{S}\otimes
\Pi _{0}^{A}\right) U_{SA}^{\dag }\right] $ which describe the states  $%
U_{SA}\left( \rho _{S}\otimes \Pi _{0}^{A}\right) U_{SA}^{\dag }$ and $%
U_{SA}\left( \sigma _{S}\otimes \Pi _{0}^{A}\right) U_{SA}^{\dag }$ undergo
an arbitrary TPCP map $\$_{SA}$ performed on the composite system S plus A.
Based on Property (III), one can easily find
\begin{equation}
\text{sgn}_{1}(\alpha )f_{\alpha }\left( \rho _{S},\delta _{S}\right) \geq 
\text{sgn}_{1}(\alpha )f_{\alpha }\left( \rho _{Sf},\sigma _{Sf}\right) .
\label{eq1r}
\end{equation}%
Suppose the TPCP map $\$_{SA}:=\left\{ \mathbb{I}_{S}\otimes \Pi
_{n}^{A}\right\} $, according to Eq. (\ref{eq}), one can replace $\rho _{Sf}$
and $\sigma _{Sf}$ in Eq. (\ref{eq1r}), respectively, by 
\begin{equation}
\rho _{Sf}\rightarrow \tilde{\rho%
}_{Sf}=\sum\limits_{n}M_{n}\rho _{S}M_{n}^{\dagger }\otimes \Pi _{n}^{A}
\end{equation}
and \begin{equation}
\sigma _{Sf}\rightarrow \tilde{\sigma}_{Sf}=\sum\limits_{n}M_{n}\sigma
_{S}M_{n}^{\dagger }\otimes \Pi _{n}^{A}.\end{equation}  Therefore, we get 
\begin{gather}
\text{sgn}_{1}(\alpha )f_{\alpha }\left( \rho _{S},\delta _{S}\right) \geq 
\text{sgn}_{1}(\alpha )f_{\alpha }\left( \tilde{\rho}_{Sf},\tilde{\sigma}%
_{Sf}\right)   \notag \\
=\text{sgn}_{1}(\alpha )\sum\limits_{n}f_{\alpha }\left( M_{n}\rho
_{S}M_{n}^{\dagger }\otimes \Pi _{n}^{A},M_{n}\sigma _{S}M_{n}^{\dagger
}\otimes \Pi _{n}^{A}\right)   \notag \\
=\text{sgn}_{1}(\alpha )\sum\limits_{n}f_{\alpha }\left( M_{n}\rho
_{S}M_{n}^{\dag },M_{n}\sigma _{S}M_{n}^{\dagger }\right)   \notag \\
=\text{sgn}_{1}(\alpha )\sum\limits_{n}p_{n}^{\alpha }q_{n}^{1-\alpha
}f_{\alpha }\left( \rho _{n},\sigma _{n}\right) ,  \label{main1}
\end{gather}%
which completes the proof.\hfill{}$\blacksquare$

Based on Lemma 1 and the preliminaries given in the previous section, we can present our main theorem as follows.

\textbf{Theorem 1.}-The coherence of a  quantum state $\rho $
can be measured by 
\begin{eqnarray}
C_{\alpha }\left( \rho \right)  &=&\min_{\delta \in \mathcal{I}}\frac{1}{%
\alpha -1}\left( f_{\alpha }^{1/\alpha }\left( \rho ,\delta \right)
-1\right)   \label{t1} \\
&=&\frac{1}{\alpha -1}\left( \sum_{j}\left\langle j\right\vert \rho ^{\alpha
}\left\vert j\right\rangle ^{1/\alpha }-1\right) ,  \label{firstd}
\end{eqnarray}%
where $\alpha \in (0,2]$, $\left\{ \left\vert j\right\rangle \right\} $ is
the reference basis and $f_{\alpha }\left( \rho ,\delta \right) =\left(
\alpha -1\right) D_{\alpha }\left( \rho ||\sigma \right) +1$ with $D_{\alpha
}\left( \rho ||\sigma \right) $ representing the Tsallis relative $\alpha $
entropy.

\textbf{Proof.}-At first, one can note that the function $x^{\alpha }$ is a
monotonically increasing function on $x$, so Eq. (\ref{firstd}) obviously
holds for positive $x$ due to Eq. (\ref{concl}).

\textit{Null.- } Since the original Tsallis entropy defined by Eq. (\ref{Tsc}) can unambiguously distinguish a coherent state from the incoherent one. Eq. (\ref{Tsc}) implies that $\sum_{j}\left\langle
j\right\vert \rho ^{\alpha }\left\vert j\right\rangle ^{1/\alpha }=1$ is
sufficient and necessary condition for incoherent states. Thus the zero $C_{\alpha }\left( \rho \right)$ is also a sufficient and necessary condition for incoherent state $\rho$. 

\textit{Convexity.-} From Ref. \cite{Liebp}, one can learn that the function $%
g(A)=$Tr$\left( XA^{p}X^{\dag }\right) ^{s}$ is convex in positive matrix $A$
for $p\in \lbrack 1,2]$ and $s\geq \frac{1}{p}$, and concave in $A$ for $p\in
(0,1]$ and $1\leq s\leq \frac{1}{p}$. Now let's assume $A=\rho $ , $%
X=\left\vert j\right\rangle \left\langle j\right\vert $ and $p=\alpha $ and $%
s=\frac{1}{\alpha }$, thus one has 
\begin{eqnarray}
g^j_\alpha(\rho)&=&\mathrm{Tr}\left( \left\vert j\right\rangle \left\langle j\right\vert\rho^{\alpha}\left\vert j\right\rangle \left\langle j\right\vert\right) ^{1/\alpha}= \left\langle j\right\vert\rho^{\alpha}\left\vert j\right\rangle  ^{1/\alpha},\end{eqnarray} 
which implies $g^j_\alpha(\rho)$ is convex in density matrix $\rho$
for $\alpha \in \lbrack 1,2]$ and $s= \frac{1}{\alpha}$, and concave in $\rho$ for $\alpha\in
(0,1]$ and $s= \frac{1}{\alpha}$. Here the subscript $\alpha$ and the superscript $j$ in $g^j_\alpha$ specifies the particular choice. So it is easy to find that 
$\frac{1}{\alpha-1}\sum_j g^j_\alpha(\rho)$ is convex for $\alpha\in (0,2]$. Considering Eq. (\ref{firstd}), one can easily show $C_{\alpha }\left( \rho \right) $
is convex in $\rho $.

\textit{Strong monotonicity.-} Now let $\{M_{n}\}$ denote the incoherent
operation, so the ensemble after the incoherent operation on the state $\rho 
$ can be given by $\left\{ p_{n},\rho _{n}\right\} $ with $p_{n}=$Tr$%
M_{n}\rho M_{n}^{\dag }$ and $\rho _{n}=M_{n}\rho M_{n}^{\dag }/p_{n}$. Thus
the average coherence $\bar{C}_{\alpha }$ is 
\begin{eqnarray}
\bar{C}_{\alpha } &=&\sum_{n}p_{n}C_{\alpha }\left( \rho _{n}\right)  \notag
\\
&=&\min_{\delta _{n}\in \mathcal{I}}\frac{1}{\alpha -1}\left(
\sum_{n}p_{n}f_{\alpha }^{1/\alpha }\left( \rho _{n},\delta _{n}\right)
-1\right) . \label{diyi}
\end{eqnarray}%
Let $\delta ^{o}$ denote the optimal incoherent state such that 
\begin{equation}
C_{\alpha}\left( \rho \right) =\frac{1}{\alpha -1}\left( f_{\alpha }^{1/\alpha
}\left( \rho ,\delta ^{o}\right) -1\right) ,
\end{equation} i.e., 
\begin{equation} 
f_{\alpha }(\rho
,\delta ^{o})=\min_{\delta \in \mathcal{I}}\mathrm{sgn}_{1}(\alpha )f_{\alpha
}(\rho ,\delta ).
\end{equation} Considering the incoherent operation $\{M_{n}\},$ we have $%
\sigma _{n}^{o}=M_{n}\delta ^{o}M_{n}^{\dagger }/q_{n}\in \mathcal{I}$ with $%
q_{n}=$Tr$M_{n}\delta ^{o}M_{n}^{\dagger }$. Therefore, one can immediately
find that%
\begin{equation}
\min_{\delta \in \mathcal{I}}\text{sgn}_{1}(\alpha )f_{\alpha }^{1/\alpha
}(\rho ,\delta )\leq \text{sgn}_{1}(\alpha )f_{\alpha }^{1/\alpha }\left(
\rho _{n},\sigma _{n}^{o}\right) ,\label{dier}
\end{equation}%
where we use the function $x^{1/\alpha }$ is monotonically increasing on $x$%
. According to Eqs. (\ref{diyi}) and (\ref{dier}), we obtain 
\begin{equation}
\bar{C}_{\alpha }\leq \frac{1}{\alpha -1}\left( \sum_{n}p_{n}f_{\alpha
}^{1/\alpha }\left( \rho _{n},\sigma _{n}^{o}\right) -1\right) .  \label{tit}
\end{equation}%
In addition, the H\"{o}lder inequality \cite{Holder} implies that for $\alpha \in (0,1),$%
\begin{equation}
\left[ \sum_{n}q_{n}\right] ^{1-\alpha }\left[ \sum_{n}p_{n}f_{\alpha
}^{1/\alpha }\left( \rho _{n},\sigma _{n}^{o}\right) \right] ^{\alpha }\geq
\sum_{n}p_{n}^{\alpha }q_{n}^{1-\alpha }f_{\alpha }\left( \rho _{n},\sigma
_{n}^{o}\right) ,
\end{equation}%
and the inequality sign is reverse for $\alpha \in (1,2],$ so Eq. (\ref{tit}%
) becomes 
\begin{eqnarray}
\bar{C}_{\alpha } &\leq &\frac{1}{\alpha -1}\left( \left[ \sum_{n}p_{n}^{%
\alpha }q_{n}^{1-\alpha }f_{\alpha }\left( \rho _{n},\sigma _{n}^{o}\right) %
\right] ^{1/\alpha }-1\right)  \notag \\
&\leq &\frac{1}{\alpha -1}\left( f_{\alpha }^{1/\alpha }\left( \rho ,\delta
^{o}\right) -1\right) =C_{\alpha },  \label{comp1}
\end{eqnarray}%
which is due to Lemma 1. Eq. (\ref{comp1}) shows the strong monotonicity of $%
C_{\alpha }.$\hfill{}$\blacksquare$

\section{Maximal coherence and several typical quantifiers}
Next, we will show that the maximal coherence can be achieved by the maximally coherent states.
At first, we assume  $\alpha\in(0,1)$. Based on the
eigen-decomposition of a $d$-dimensional state $\rho :\rho =\sum\limits_{k}\lambda _{k}\left\vert
\psi _{k}\right\rangle \left\langle \psi _{k}\right\vert $ with $\lambda
_{k} $ and $\left\vert \psi _{k}\right\rangle $ representing the eigenvalue
and eigenvectors, we have 
\begin{eqnarray}
\sum_{j}\left\langle j\right\vert \rho ^{\alpha }\left\vert j\right\rangle
^{1/\alpha } &=&\sum_{j}\left( \sum_{k}\lambda _{k}^{\alpha }\left\vert
\left\langle \psi _{k}\right. \left\vert j\right\rangle \right\vert
^{2}\right) ^{1/\alpha }  \notag \\
&\geq &d\left( \sum_{jk}\frac{\lambda _{k}^{\alpha }}{d}\left\vert
\left\langle \psi _{k}\right. \left\vert j\right\rangle \right\vert
^{2}\right) ^{1/\alpha }  \notag \\
&\geq &d\left( \sum_{k}\frac{\lambda _{k}^{\alpha }}{d}\right) ^{1/\alpha
}\geq d^{\frac{\alpha -1}{\alpha }}.  \label{lb}
\end{eqnarray}%
One can easily find that the lower bound Eq. (\ref{lb}) can be attained by the maximally coherent states $\rho_m=\left\vert \Psi\right\rangle\left\langle\Psi\right\vert$ with $\left\vert\Psi\right\rangle=\frac{1}{\sqrt{d}}\sum_je^{i\phi_j}\left\vert j\right\rangle$. Correspondingly, the coherence is given by \begin{equation}C_{0<\alpha<1}(\rho_m)=\frac{1}{1-\alpha}(1-d^{\frac{\alpha -1}{\alpha }}).\end{equation} Similarly, for  $\alpha \in (1,2]$,   the function $x^{1/\alpha }$ is
concave, which leads to that Eq. (\ref{lb}) with the
inverse inequality sign holds. The inequality can also saturate for $\rho_m$. The corresponding coherence is given by \begin{equation}C_{1<\alpha\leq2}(\rho_m)=\frac{1}{\alpha-1}(d^{\frac{\alpha -1}{\alpha }}-1).\end{equation}

 $C_{\alpha }\left( \rho \right) $ actually defines a family of
coherence measures related to the Tsallis relative $\alpha $ entropy. This
family includes several typical coherence measures. As mentioned above, the
most prominent coherence measure belonging to this family is the coherence
in terms of relative entropy, i.e., $C_{1}\left( \rho \right) =S(\rho )$.

One can also find that 
\begin{eqnarray}
C_{1/2}\left( \rho \right)  &=&\min_{\delta \in \mathcal{I}}2\left( 1-\left[
Tr\sqrt{\rho }\sqrt{\delta }\right] ^{2}\right)   \notag \\
&=&\min_{\delta \in \mathcal{I}}\left\Vert \sqrt{\rho }-\sqrt{\delta }%
\right\Vert _{2}^{2}  \notag \\
&=&1-\sum_{i}\left\langle i\right\vert \sqrt{\rho }\left\vert i\right\rangle
^{2}
\end{eqnarray}%
with $\left\Vert \cdot \right\Vert _{2}$ denoting $l_{2}$ norm. So the  $%
l_{2}$ norm has been revived for coherence measure by considering the square
root of the density matrices. This is much like the quantification of
quantum correlation proposed in Ref. \cite{luoq}. In addition,  $C_{1/2}(\rho)$ can also
be rewritten as
\begin{equation}
C_{1/2}\left( \rho \right)  =-\frac{1}{2}\sum_i\mathrm{Tr}\left\{\left [\sqrt{\rho},\left\vert i\right\rangle\left\langle i\right\vert\right]^2\right\}
\end{equation}
which is just the coherence measure based on the skew information \cite{skew1,skew2}.

Finally, one can also see
that 
\begin{eqnarray}
C_{2}\left( \rho \right)  &=&\min_{\delta \in \mathcal{I}}\left( \sqrt{%
Tr\rho ^{2}\delta ^{-1}}-1\right)   \notag \\
&=&\sum_{i}\left\langle i\right\vert \rho ^{2}\left\vert i\right\rangle
^{1/2}-1
\end{eqnarray}%
which is a simple function of the density matrix.

\section{Discussions and conclusion}
We establish a family of coherence measures that are closely related to the Tsallis relative $\alpha$ entropy.
We prove that these coherence measures satisfy all the required criteria for a satisfactory coherence measure especially including 
the strong monotonicity.
 We also 
show this family of coherence measures includes several typical coherence measures such as the coherences measure based on von Neumann entropy, skew information and so on. Additionally, we show how to validate the $l_2$ norm as a coherence measure. Finally, we would like to emphasize that the convexity and the strong
monotonicity could be two key points which couldn't easily be compatible with each other to some extent. Fortunately, Ref. \cite{Liebp} provides the important knowledge to harmonize both points in this paper. This work builds the bridge between the Tsallis relative $\alpha$ entropy and the strong monotonicity and provides the important alternative quantifiers for the  coherence quantification. This could shed new light on the strong monotonicity of other candidates for coherence measure.

\section{Acknowledgements}
 This work was supported by the National Natural Science
Foundation of China, under Grant No.11375036, the Xinghai Scholar
Cultivation Plan and the Fundamental Research Funds for the Central
Universities under Grant No. DUT15LK35 and No. DUT15TD47.


\begin{thebibliography}{99}




\bibitem{Engel} G. S. Engel, T. R. Calhoun, E. L. Read, T.-K. Ahn, T. Man%
\v{c}al, Y.-C. Cheng, R. E. Blankenship, and G. R. Fleming, Nature (London) 
\textbf{446}, 782 (2007).

\bibitem{Plenio} M. B. Plenio, and S. F. Huelga, New J. Phys. \textbf{10},
113019 (2008).

\bibitem{Coll} E. Collini, C. Y. Wong, K. E. Wilk, P. M. G. Curmi, P.
Brumer, and G.D. Scholes, Nature (London) \textbf{463}, 644 (2010).

\bibitem{loyd} S. Lloyd, J. Phys. Conf. Ser. \textbf{302}, 012037 (2011).

\bibitem{licm} C. M. Li, N. Lambert, Y.-N. Chen, G. Y. Chen, and F. Nori,
Sci. Rep. \textbf{2}, 885 (2012).

\bibitem{Huel} S. Huelga, and M. Plenio, Contemp. Phys. \textbf{54}, 181
(2013).

\bibitem{Ryb} L. Rybak, S. Amaran, L. Levin, M. Tomza, R. Moszynski, R.
Kosloff, C. P. Koch, and Z. Amitay, Phys. Rev. Lett. \textbf{107}, 273001
(2011).

\bibitem{reb} P. Rebentrost, M. Mohseni, and A. Aspuru-Guzik, J. Phys. Chem.
B \textbf{113}, 9942 (2009).

\bibitem{wit} B. Witt, and F. Mintert, New J. Phys. 15, 093020 (2013).

\bibitem{berg} J. \AA berg, Phys. Rev. Lett. \textbf{113}, 150402 (2014).

\bibitem{Nar} V. Narasimhachar, and G. Gour, arXiv: 1409.7740 [quant-ph].

\bibitem{Horo} P. \'{C}wikli\'{n}ski, M. Studzi\'{n}ski, M. Horodecki, and
J. Oppenheim, arXiv: 1405.5029 [quant-ph].

\bibitem{Los1} M. Lostaglio, D. Jennings, and T. Rudolph, Nat. Commun. 
\textbf{6}, 6383 (2015).

\bibitem{Los2} M. Lostaglio, K. Korzekwa, D. Jennings, and T. Rudolph, Phys.
Rev. X \textbf{5}, 021001 (2015).

\bibitem{Glauber} R. J. Glauber, Phys. Rev. \textbf{131}, 2766 (1963).

\bibitem{Sudarshan} E. C. G. Sudarshan, Phys. Rev. Lett. \textbf{10}, 277 (1963).

\bibitem{Scully} M. O. Scully and M. S. Zubairy, \textit{Quantum Optics} (Cambridge University Press, Cambridge, England, 1997).




\bibitem{Pleniom} T. Baumgratz, M. Cramer, and M. B. Plenio, Phys. Rev.
Lett. \textbf{113}, 140401 (2014).



\bibitem{Lewen} S. Rana, P. Parashar, and M. Lewenstein, Phys. Rev. A 
\textbf{93}, 012110 (2016).

\bibitem{Giro} D. Girolami, Phys. Rev. Lett. \textbf{113}, 170401 (2014).

\bibitem{Napoli} C. Napoli, T. R. Bromley, M. Cianciaruso, M. Piani, N.
Johnston, and G. Adesso, Phys. Rev. Lett. \textbf{116}, 150502 (2016).

\bibitem{Rast} A. E. Rastegin, Phys. Rev. A \textbf{93}, 032136 (2016).



\bibitem{Piani} M. Piani, M. Cianciaruso, T. R. Bromley, C. Napoli, N.
Johnston, and G. Adesso, Phys. Rev. A \textbf{93}, 042107 (2016).

\bibitem{Winter} A. Winter, and D. Yang, Phys. Rev. Lett. \textbf{116},
120404 (2016).

\bibitem{Du} S. Du, Z. Bai, and Y. Guo,  Phys. Rev.
A \textbf{91}, 052120 (2015).

\bibitem{Chi} E. Chitambar, A. Streltsov, S. Rana, M. N. Bera, G. Adesso,
and M. Lewenstein, Phys. Rev. Lett. \textbf{116}, 070402 (2016).

\bibitem{Chi2} E. Chitambar, and M.-H. Hsieh, Phys. Rev. Lett. \textbf{117},
020402 (2016).

\bibitem{Chi3} E. Chitambar, and Gilad Gour, Phys. Rev. Lett. \textbf{117},
030401 (2016).

\bibitem{Radha} C. Radhakrishnan, M. Parthasarathy, S. Jambulingam, and T.
Byrnes, Phys. Rev. Lett. \textbf{116}, 150504 (2016).


\bibitem{Marvian} I. Marvian, and R. W. Spekkens, Phys. Rev. A \textbf{90},
062110 (2014).

\bibitem{Marvian2} I. Marvian, R. W. Spekkens, and P. Zanardi, Phys. Rev. A 
\textbf{93}, 052331 (2016).

\bibitem{Yao} Y. Yao, X. Xiao, L. Ge, and C. P. Sun, Phys. Rev. A \textbf{92}%
, 022112 (2015).

\bibitem{Sing} U. Singh, L. Zhang, and A. K. Pati, Phys. Rev. A \textbf{93},
032125 (2016).


\bibitem{Yu09} C. S. Yu, and H. S. Song, Phys. Rev. A \textbf{80}, 022324
(2009).
\bibitem{Stre} A. Streltsov, U. Singh, H. S. Dhar, M. N. Bera, and G.
Adesso, Phys. Rev. Lett. \textbf{115}, 020403 (2015).

\bibitem{Ma} J. Ma, B. Yadin, D. Girolami, V. Vedral, and M. Gu, Phys. Rev.
Lett. \textbf{116}, 160407 (2016).

\bibitem{Tan} K. C. Tan, H. Kwon, C. Y. Park, and H. Jeong, Phys. Rev.
A \textbf{94}, 022329 (2016).



\bibitem{Fan} L. H. Shao, Z. J. Xi, H. Fan, and Y. M. Li, Phys. Rev.
A \textbf{91}, 042120 (2015).

\bibitem{lisa} L. Borland, A. R. Plastino, and C. Tsallis, J. Math. Phys. \textbf{39}, 6490 (1998).

\bibitem{Tsallis} C. Tsallis, et al., in \textit{Nonextensive Statistical Mechanics and Its Applications}, edited by S. Abe and Y. Okamoto (Springer-Verlag, Heidelberg, 2001).

\bibitem{Gour} F. G. S. L. Brand\~{a}o, and Gilad Gour, Phys. Rev. Lett. \textbf{115},
070503 (2015).



\bibitem{Hiai} F. Hiai, M. Mosonyi, D. Petz, and C. B\`{e}ny, Rev. Math. Phys.
\textbf{23}, 691 (2011).



\bibitem{Nielsen} M. A. Nielsen, and I. L. Chuang, \textit{Quantum computation an quantum information} 
(Cambridge University Press, Cambridge, England, 2000).


\bibitem{Liebp} E. A. Carlen, and E. H. Lieb, Lett. Math. Phys. \textbf{83}, 107 (2008).


\bibitem{Holder} J. C. Kuang, \textit{Applied inequalities} (Shandong Science and Technology Press, Jinan, China, 2012).

\bibitem{luoq} L. N. Chang, and S. L. Luo, Phys. Rev. A \textbf{87}, 062303 (2013).

 \bibitem{skew1} E. P. Wigner, and M. M. Yanase, Proc. Natl. Acad. Sci. 49,
910 (1963).

\bibitem{skew2} E. H. Lieb, Adv. Math. \textbf{11}, 267 (1973).
 
\end{thebibliography}
\end{document}